\documentclass{desyprocA4}

\usepackage{amsmath,subfig,cite}
\usepackage[bottom]{footmisc}

\newcommand{\gsm}{\Gamma^\text{SM}}
\newcommand{\ghid}{\Gamma^\text{hid}}
\newcommand{\gi}{\Gamma^\text{inv}}
\newcommand{\gv}{\Gamma^\text{vis}}
\newcommand{\ghh}{\Gamma^{HH}}

\newcommand{\bri}{\text{BR}^\text{inv}}

\newcommand{\gev}{\text{GeV}}

\newcommand{\ifb}{\text{fb}^{-1}}

\newcommand{\gl}[1]{(\ref{#1})}

\begin{document}

\title{The Higgs Portal from LHC to ILC}

\author{
  {\slshape Christoph Englert} \\[1ex]
  Institute for Particle Physics
  Phenomenology, Department of Physics, Durham University, United
  Kingdom\\
}

\contribID{xy}

\confID{1964}  
\desyproc{DESY-PROC-2010-01}
\acronym{PLHC2010} 
\doi  

\maketitle

\begin{abstract}
  Interpretations of searches for the Higgs boson are governed by
  model-dependent combinations of Higgs production cross sections and
  Higgs branching ratios.  
  Mixing of the Higgs doublet with a hidden sector captures
  modifications from the Standard Model Higgs phenomenology in the
  standard search channels in a representative way, in particular
  because invisible Higgs decay modes open up.
  As a consequence, LHC exclusion bounds, which disfavor a heavy
  Standard Model Higgs can be consistently understood in terms of a
  standard-hidden mixed Higgs system.
  Shedding light on the possible existence of such an admixture with a
  hidden sector and quantifying the resemblance of an eventually
  discovered scalar resonance with the Standard Model Higgs crucially
  depends on measurement of invisible decays. This task will already
  be tackled at LHC, but eventually requires the clean environment of
  a future linear collider to be ultimately completed.
\end{abstract}

\section{Introduction}
Recent measurements at the CERN Large Hadron Collider
\cite{recent,HiggsATLAS,HiggsCMS,Rol} constrain a SM-like Higgs to be
lighter than $m_H\lesssim 130~{\text{GeV}}$ at 95\% confidence
level. Moreover, both ATLAS and CMS have observed an excess for Higgs
masses around 125 GeV, consistent with each other. These tantalizing
hints for a light Higgs boson in the multilepton $H\to 4 \ell$ and,
more importantly, in the $H\to \gamma \gamma$ channels are in
excellent agreement with theoretical expectations, which have been
coined by electroweak precision measurements performed during the LEP
era \cite{Alcaraz:2006mx}.

The accumulated statistics of approximately 5 fb$^{-1}$ per
experiment, however, is yet too small to draw a conclusive picture
about mechanism of electroweak symmetry breaking. Since the assumption
of SM-like production and decay explicitly enter the hypothesis tests
that lead to the formulation of the LHC exclusion limits, the
quantitative resemblance of the observed phenomenology with the SM is,
in fact, not entirely transparent. Instead of mere numerical agreement
of data with the SM Higgs hypothesis, we can understand the exclusion
limits as a measure of how much a more general theory is bound to
coincide with the SM in the light of current experimental
observations. This exercise naturally yields model-dependent
statements, but there is only a limited number of phenomenological
patterns of how the Higgs can evade detection\footnote{Note that in
  non-local theories of electroweak symmetry breaking the Higgs can be
  significantly underproduced \cite{unhiggs}.}~\cite{Englert:2011us}.
The extension of the Higgs sector by including invisible decay modes
and constrain them by measurements is crucial for the
re-interpretation of the exclusion bounds is this context. A
substantial non-zero branching ratio would signalize a non-standard
Higgs sector while being in perfect agreement with a non-observation
of the Higgs at the moment.

Constraining invisible branching ratios is a difficult and challenging
task at hadron colliders with their busy final states
\cite{DeRoeck:2009id}. A statistically significant determination of an
invisible Higgs branching ratio requires large statistics (if possible
after all) as experimental systematics set the scale of uncertainty.
Systematics vastly improves when studying the Higgs sector at a future
linear collider. There, $e^+e^-\to HZ$ associated production provides
an extremely clean laboratory process to study invisible decays in a
model-independent way in recoil analyses \cite{schumacher}.  At the
LHC, only ratios of branching fractions are accessible in a
model-independent fashion, but absolute branching ratio predictions can
be formulated in specified models \cite{Bock:2010nz}. Hence,
performing such an analysis at a future linear collider is going to be
of utmost importance to study the Higgs boson in full detail after its
discovery at the LHC.

\begin{figure}[!t]
  \begin{center}
    \includegraphics[height=0.33\textwidth]{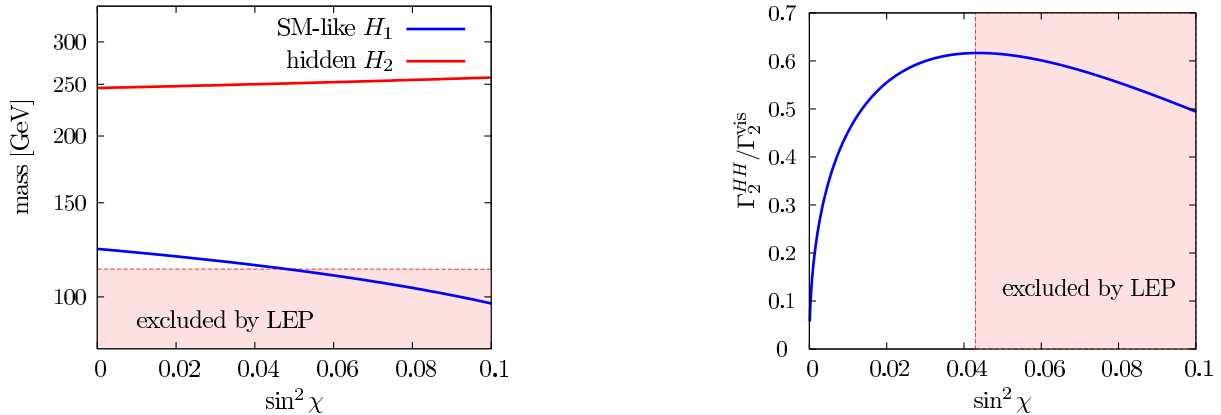}
    \caption{\label{fig:Masses12} Left: masses of the light SM-like
      Higgs boson $H_1$ (blue) and the heavy Higgs boson $H_2$
      (red). We choose the vacuum expectation values $v_h = v_s = 246$
      GeV and $\lambda_s = \lambda_h /4 = 1/8$ for illustration
      purposes. The shaded region displays the LEP
      bound~\cite{Alcaraz:2006mx}. Right: cascade decay width $\ghh_2$
      as a function of $\sin^2\chi\,$ for the same parameters. Again,
      the region in which $H_1$ is excluded by LEP is shaded. The figures
      are taken from Ref.~\cite{Englert:2011yb}.}
  \end{center}
\end{figure}

\section{The Higgs portal}
We introduce invisible decay channels in an efficient and
theoretically consistent way via a particular type of hidden valley
\cite{hiddenvalley} interaction in the Higgs sector. The SM Higgs
doublet $\phi_s$ is coupled to a hidden sector scalar field $\phi_h$
via the gauge-invariant and renormalizable operator $ |\phi_s|^2
|\phi_h|^2$ so that the potential reads \cite{Higgs.portal}
\begin{equation}
 \label{eq:potential}
 \mathcal{V} =
 \mu^2_s |\phi_s|^2 + \lambda_s |\phi_s|^4 
 \; + \;
 \mu^2_h |\phi_h|^2 + \lambda_h |\phi_h|^4
 \; + \;
 \eta_\chi |\phi_s|^2 |\phi_h|^2 \, .
\end{equation}
The mass parameters $\mu_j$ can be substituted by $v_j$ after
expanding the two Higgs fields about their vacuum expectation values
$v_j^2 = (- \mu_j^2 - \eta_\chi v_i^2 /2)/\lambda_j$ ($i \neq j =
s,h$). The electroweak gauge boson masses are generated exclusively by
the visible fields' vacuum expectation. The so-called Higgs portal
interaction operator $\sim \eta$ rotates $s,h$ states into the mass
eigenstates
\begin{equation}
  \begin{split}
    H_1 &=  \cos\chi \, H_s + \sin\chi \, H_h \\
    H_2 &= -\sin\chi \, H_s + \cos\chi \, H_h \,,
    \label{eq:mixi}
  \end{split}
\end{equation}
where $\sin\chi$ is the characteristic mixing angle, which affects the
production cross sections $\sigma_{1,2}$ and visible and invisible
decay widths $\Gamma_{1,2}^{\rm{vis,inv}}$ of the two Higgs bosons
in an universal fashion \cite{Bock:2010nz}
\begin{subequations}
\label{eq:parameters}
\begin{equation}
  \sigma_{1,2}  = \cos^2\chi \, \{\sin^2\chi\} \, \sigma^\text{SM}_{1,2}
\end{equation}
and
\begin{alignat}{5}
  \gv_1  =& \cos^2\chi \, \gsm_1 \qquad \text{and} \qquad 
& \gv_2  =& \sin^2\chi \, \gsm_2                            \notag \\
  \gi_1  =& \sin^2\chi \, \ghid_1 \qquad\; \text{and} \qquad
& \gi_2  =& \cos^2\chi \, \ghid_2                           \,.
\end{alignat}
\end{subequations}
The index ``SM'' refers to the values in the SM, and the information
on the hidden sector is encoded in the ``hid'' quantities.  If
kinematically allowed, {\it{i.e.}} for $m_{H_2} \gtrsim 2m_{H_1}$ we
can have additional cascade decays (in the following we take $H_1$ to
be the lighter, mostly SM-like state by definition), which, depending
on the combinations of the fundamental parameters, can play a
significant role \cite{Englert:2011yb}. We exemplarily show a Higgs
spectrum as a function of $\sin^2\chi$ in Fig.~\ref{fig:Masses12}. The
relations between the suppression factors $\sin^2\chi$, the masses
$m_{H_1},m_{H_2}$ and the fundamental lagrangian parameters of
Eq.~\gl{eq:potential} can be obtained by straightforward calculation
and we refer the reader to Refs.~\cite{Bock:2010nz,Englert:2011yb} for
further details.

\begin{figure}[!b]
  \begin{center}
    \includegraphics[height=0.7\textwidth]{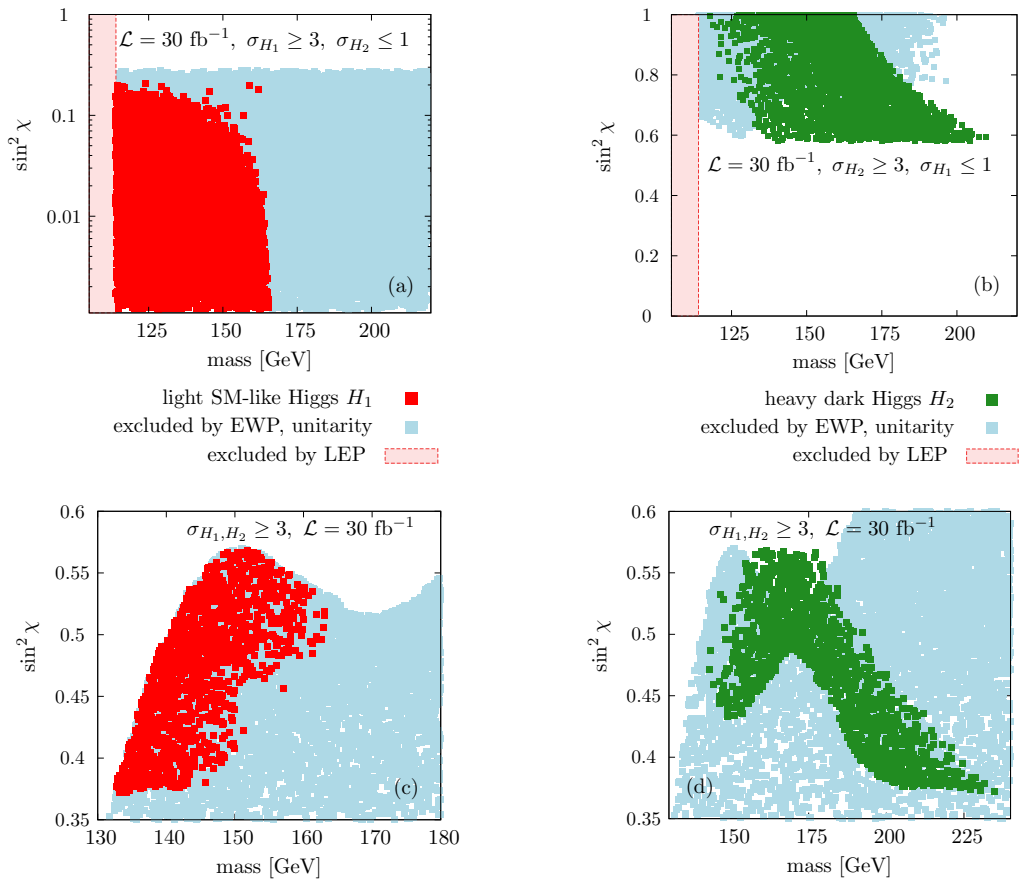}
    \caption{\label{fig:Mx} Scan over the Higgs portal model
      Eq.~\gl{eq:potential} for parameter ranges
      $v_h\in(0~\text{GeV},246~\text{GeV}], v_s= 246~\text{GeV},
      \lambda_h\in(0,4\pi], \lambda_s\in(0,4\pi]$, and
      $\eta_\chi\in[-4\pi,4\pi]$. The hidden Higgs decay width is
      identified with the SM decay width for demonstration purposes,
      {\it i.e.}  $\ghid \equiv \gsm$. LEP constraints and bounds from
      $S,T,U$~\cite{Alcaraz:2006mx} and unitarity are
      included. Panel~(a) displays the sensitivity for $H_1$ only,
      panel~(b) for $H_2$ only, and panels~(c) and (d) show where the
      LHC is sensitive to both $H_1$ and $H_2$ at the same time for
      $30~\ifb$ at $\sqrt{s}=14$~TeV. The figures are taken from
      Ref.~\cite{Englert:2011yb}.}
  \end{center}
\end{figure}

The model of Eq.~\gl{eq:potential} is subject to constraints by
electroweak precision observables and partial wave unitarity. A
guiding principle toward the validity of a model is the comparison of
the model's prediction of the Peskin-Takeuchi parameters
\cite{Peskin:1990zt} with measurements performed at LEP
\cite{Alcaraz:2006mx}. These give rise to the strongest constraints on
the Higgs portal model\footnote{For a discussion of perturbativity and
  stability of the potential Eq.~\gl{eq:potential} see
  Ref.~\cite{Bowen:2007ia}.}. This is easy to understand: for larger
mixing angles $\sin^2\chi \to 1$ we effectively deal with a heavy
Higgs model which is tightly constraint\footnote{Note that in a
  realistic scenario we can expect kinetic mixing with a heavy $U(1)$
  boson \cite{Bowen:2007ia}, which again loosens the electroweak
  precision constraints.} by the measurements of
\cite{Alcaraz:2006mx}. At the same time, the isometry Eq.~\gl{eq:mixi}
restores unitarity in the high energy limit.

\section{Higgs portal lessons from the LHC}

Altogether the model predicts the four different phenomenological
scenarios of Tab.~\ref{tab:tab1} for standard Higgs resonance searches
at typical LHC Higgs discovery luminosities ($\sqrt{s}=14$ TeV),
{\emph{cf.}}  Fig.~\ref{fig:Mx}, where we assume
$\Gamma^{\rm{hid}}=\Gamma^{\rm{SM}}$ for simplicity. Apart from a
small window in $\sin^2\chi$, the Higgs portal can be explored in its
most symmetric version already at typical SM Higgs discovery
luminosities~\cite{AtlasTDR}.

\begin{figure}[!b]
  \begin{center}
    \subfloat[][\label{fig:mud1}]{
    \includegraphics[width=0.48\textwidth]{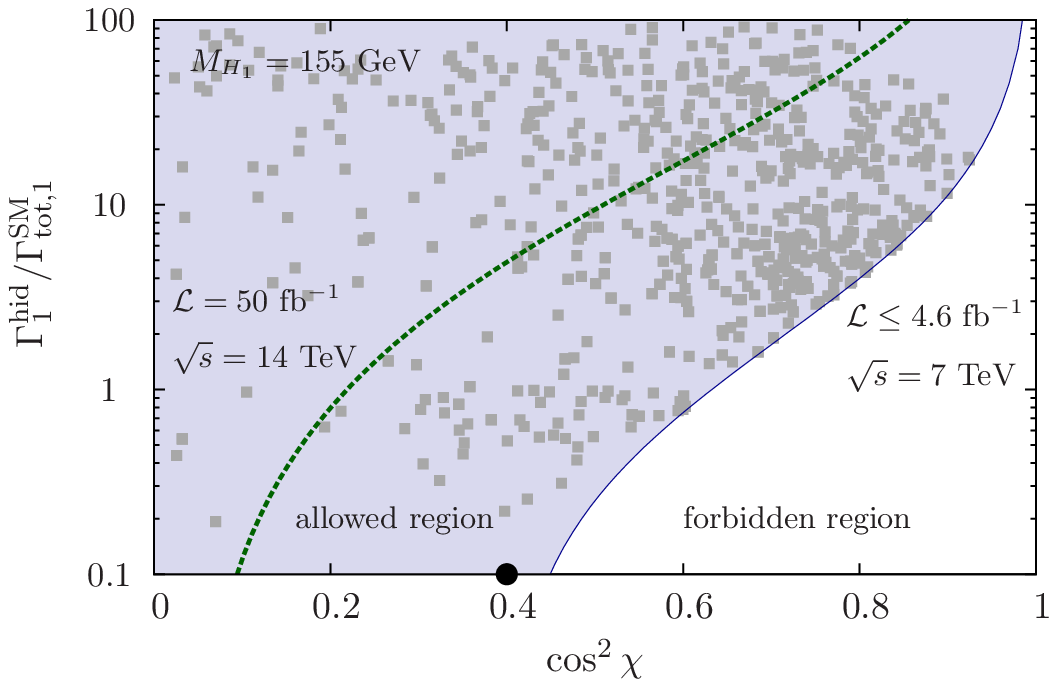}}
    \hfill
    \subfloat[][\label{fig:mud2}]{
    \includegraphics[width=0.48\textwidth]{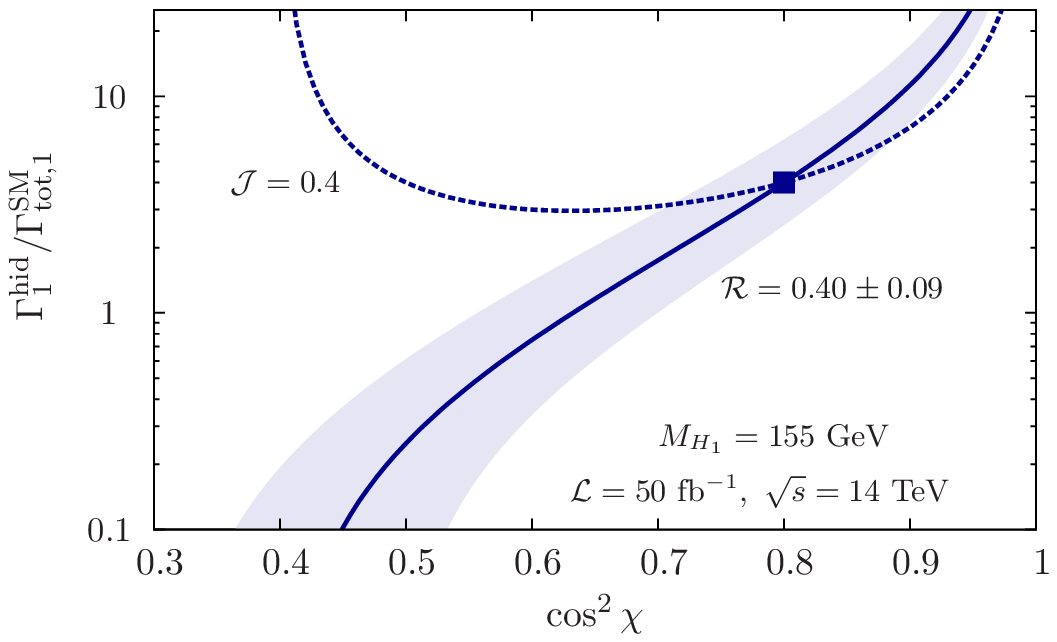}
    }
    \caption{\label{fig155} Left: bounds on the mixing and hidden
      decay width of $H_1$ for the point $M_{H_1} = 155$ $\text{GeV};
      \mathcal{R} = 0.4$ in the standard-hidden Higgs scenario, based
      on current experimental
      results~\cite{HiggsATLAS,HiggsCMS,Rol}. The regions dappled by
      small squares are compatible with unitarity and precision
      measurements. The dot indicates the $\Gamma_1^\text{hid} \to 0$
      limit of the exclusion curve at $\mathcal{R}$. The dotted line
      indicates the projected search limit for $\mathcal{L} =
      50~\ifb$. Right: bounds due to hidden Higgs searches at the LHC
      for established Higgs masses and cross sections. The figures are
      taken from Ref.~\cite{Englert:2011aa}.}
  \end{center}
\end{figure}
%
\begin{table}[t] 
  \begin{center}
    \begin{tabular}{c c r l}
      \hline
      & LHC sensitivity after 30~$\ifb$ to &\\
      \hline
      $\sin^2 \chi\lesssim 0.2$    &  only $H_1$  & ($\sigma_{H_1}\geq 3,~\sigma_{H_2}\leq 1$)        \\
      $0.3 \lesssim \sin^2 \chi \lesssim 0.4$ &  neither $H_1$ nor $H_2$ & ($\sigma_{H_1,H_2}< 3$) \\
      $ 0.4\lesssim \sin^2 \chi \lesssim 0.6$  & both $H_1$ and $H_2$ &  ($\sigma_{H_1,H_2}\geq 3$) \\
      $\sin^2\chi \gtrsim 0.6$          &  only $H_2$ & ($\sigma_{H_1}\leq 1,~\sigma_{H_2}\geq 3$)      \\ 
      \hline
    \end{tabular} 
    \caption{\label{tab:tab1} Result of Higgs searches at the LHC
      ($\sqrt{s}=14$~TeV) with a luminosity of 30~$\ifb$, $\sigma$
      refers to the sensitivity in terms of
      signal/$\sqrt{\text{background}}$.}
  \end{center}
\end{table}

Relaxing the assumption $\Gamma^{\rm{hid}}=\Gamma^{\rm{SM}}$ changes
the picture. In fact, there is good reason to also consider the
situation $\Gamma^{\rm{hid}}\gg\Gamma^{\rm{SM}}$, since the hidden
decay width parametrizes our lack of knowledge about the dynamics in
the hidden sector, which can be strong. To study the implications for
general $\Gamma^{\rm{hid}}/\Gamma^{\rm{SM}}$ choices we examine the
the 95\% confidence level bounds which are formulated by the LHC
collaborations with respect to the SM cross section. In the portal
model of Eq.~\gl{eq:potential},\gl{eq:mixi} these can be expressed
as~\cite{Englert:2011aa}
\begin{equation}
  \dfrac{\sigma[pp \to H_1 \to F]}{\sigma[pp \to H_1 \to F]^\text{SM}} 
  = \dfrac{\cos^2\chi}{1 + \tan^2\chi \, [{\Gamma^\text{hid}_1}/{\Gamma^\text{SM}_{\text{tot},1}}]}
  \leq \mathcal{R}                                                                            \,,
\label{eq:excl}
\end{equation}
where $\cal{R}$ denotes the observed exclusion limit. An identical
quantity can be derived from future constraints on invisible decays
\cite{Englert:2011us,Englert:2011aa,DeRoeck:2009id}:
\begin{equation}
  \dfrac{\sigma[pp \to H_1 \to inv]}{\sigma[pp \to H_1]^\text{SM}}
  = \dfrac{\sin^2\chi \, [\Gamma^\text{hid}_1 / \Gamma^\text{SM}_{\text{tot},1}]}
  {1 + \tan^2\chi \, [{\Gamma^\text{hid}_1}/{\Gamma^\text{SM}_{\text{tot},1}}]}
  \leq \mathcal{J}                                                            \,.
\label{eq:inv}
\end{equation}

In Fig.~\ref{fig155} we exemplarily examine the implications of the
current Higgs exclusion bounds for $m_H=155$ GeV in the
$\Gamma_1^{\rm{hid}}/\Gamma_1^{\rm{SM}}$-$\cos^2\chi$ plane. For this
particular Higgs mass the experiments observe $R=0.4$
\cite{HiggsATLAS,HiggsCMS,Rol}. From Fig.~\ref{fig155} we learn that
there is a variety portal parameter choices which can accommodate the
current phenomenological findings.

An additional constraint can be imposed in the same plane by
constraining invisible decays at the LHC\footnote{Such an analysis has
  not been performed by the experiments, but exiting analyses were
  adopted in Ref.~\cite{Englert:2011us,wbfinv}, demonstrating
  potentially sufficient sensitivity to ${\cal{J}}\sim 1$ for the
  combined 2011 data set.}. Typically this involves large statistics
when the \cite{Englert:2011aa} visible cross section of the $H_1$
state is already measured, {\it{i.e.}}  the inequality of
Eq.~\gl{eq:excl} becomes an equality within the uncertainty given by
statistics and systematics. If ${\cal{J}}$ is yet to be understood as
a 95\% confidence level exclusion \cite{DeRoeck:2009id}, we do not
have the enough information to reconstruct the all parameters of
Eq.~\gl{eq:parameters}.

\begin{figure}[!t]
  \begin{center}
    \includegraphics[width=0.9\textwidth]{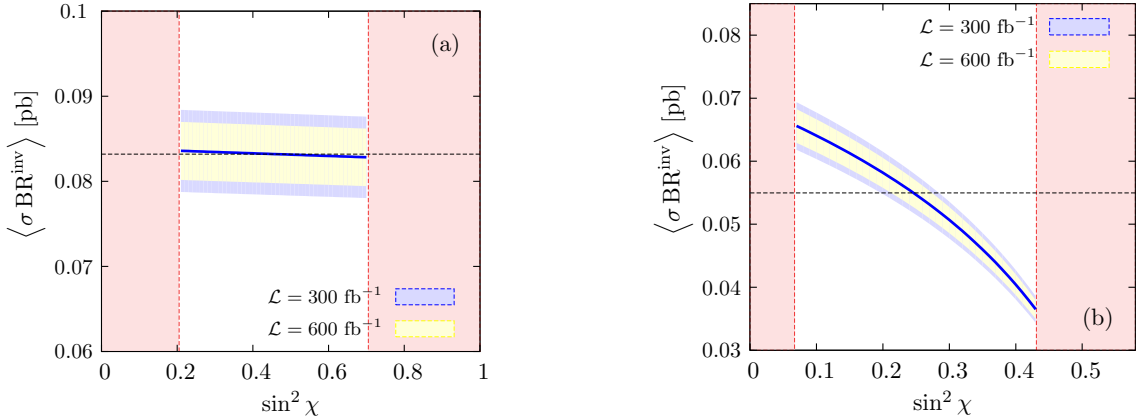}
    \caption{\label{fig3} Reconstruction of the mixing angle from a
      measurement of the superposition of the invisible decays. The
      shaded area is theoretically not allowed due to positivity of
      the cross section ratio ${\cal{R}}_i$, uncertainties of these
      parameters are not considered. (a) is a degenerate mass spectrum
      $M_{H_1}=140~\gev,~M_{H_2}=160~\gev$, (b) is a mass spectrum
      $M_{H_1}=115~\gev,~M_{H_2}=300~\gev$, where the mixing can in
      principle be reconstructed due to Eq.~\gl{eq:disentangle} and
      comments below. Uncertainties follow from statistics
      only. The figures are taken from Ref.~\cite{Englert:2011yb}.}
  \end{center}
\end{figure}

In fact, when comparing to the SM Higgs potential, the multitude of
observables which are potentially accessible in addition to the SM,
{\emph{i.e.}} the Higgs resonance masses and the cascade decay width
if present, allow for a {\emph{full}} reconstruction strategy of the
Higgs portal potential Eq.~\gl{eq:potential}. An absolutely crucial
input for this analysis is the {\emph{measurement}} of
${\cal{J}}$. The measurement of ${\cal{R}}$ for both Higgs states will
eventually be possible at the LHC for the bulk of the parameter
space. The measurement of ${\cal{J}}$ at the LHC, however, is limited
by systematics \cite{DeRoeck:2009id} and the fact that we measure a
superposition of invisible rates of the two Higgs states (on top of a
challenging background) at hadron colliders
\begin{equation}
  \label{eq:disentangle}
  \langle \sigma \bri \rangle \sim f(\Lambda) - [\cos^2 \chi + 
  \{\sigma^\text{SM}_2 / \sigma^\text{SM}_1\} \, \sin^2 \chi] \, ,
\end{equation}
where $f(\Lambda)$ depends on the trilinear coupling (if accessible)
in the invisible cascade decay $H_2\to H_1H_1\to \rm{invisible}$. Even
if a measurement turns out to possible, we rely on the separation of
the two Higgs states to lift the degeneracy in the invisible decay
channel ({\emph{cf.}} Fig.~\ref{fig3}). More concretely, in order to
project out the $\cos^2\chi$ component in Eq.~\gl{eq:disentangle} we
need $\sigma^\text{SM}_2 / \sigma^\text{SM}_1\ll 1$, {\emph{i.e.}}
$m_{H_2}\gg m_{H_1}$, unless we have a significant trilinear coupling
in the resolved cascade decay, which can be used to constrain the
mixing parameters.

In total the LHC can not cover the entire parameter space of the Higgs
portal model Eq.~\gl{eq:potential}.

\begin{figure}[!t]
  \begin{center}
    \subfloat[][\label{fig:mud1b}]{
      \includegraphics[width=0.48\textwidth]{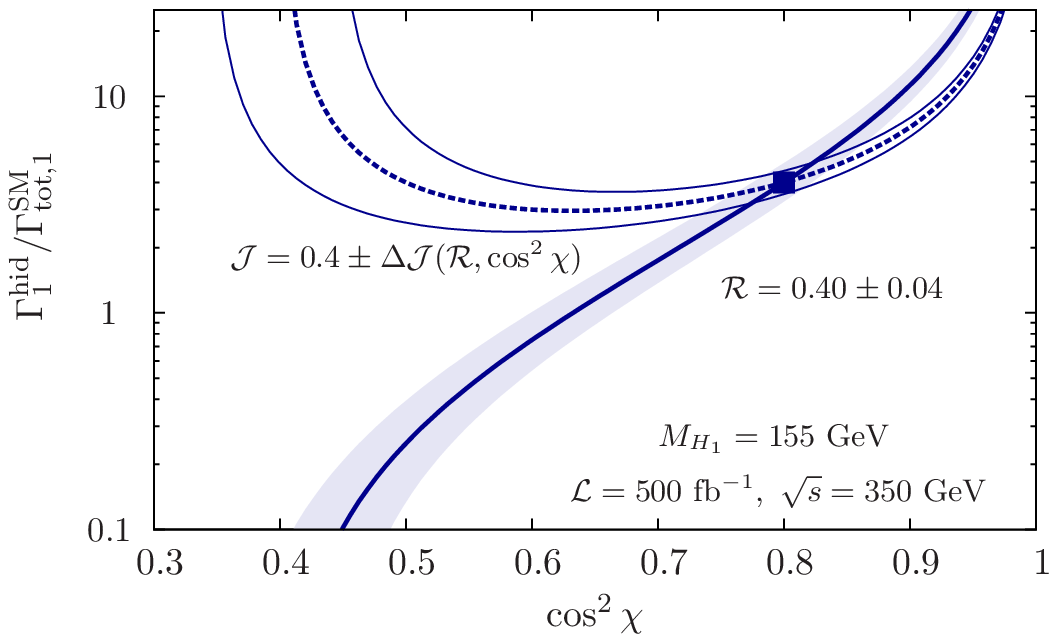}
      }\hfill
      \subfloat[][\label{fig:mud2b}]{
      \includegraphics[width=0.48\textwidth]{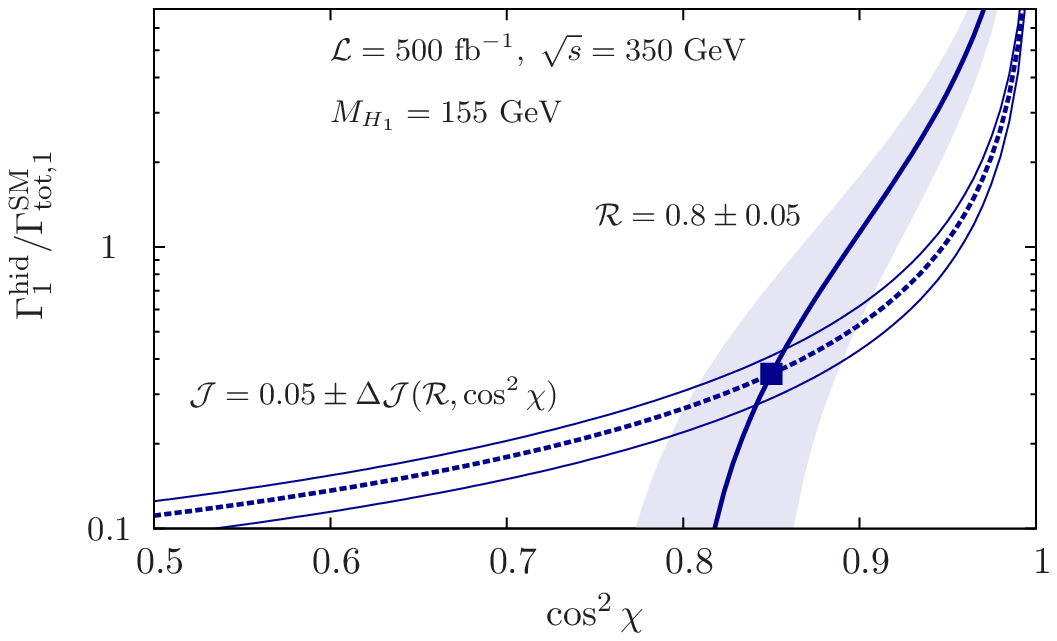}
      }
      \caption{\label{fig4} Left: the scenario of Fig~{\ref{fig:mud2}}
        at a linear collider ($\sqrt{s}=500~\gev$,
        ${\cal{L}}=500~\ifb$). Right: a Higgs portal scenario with
        small ${\cal{J}}$. The uncertainties are adopted from
        Refs.~\cite{schumacher,teslatdr}, the figure are taken from
        Ref.~\cite{Englert:2011aa}.}
  \end{center}
\end{figure}

\section{Higgs Portal spectroscopy at a linear collider}
The systematics-plagued determination of invisible branching ratio of
the individual resonances can be cured at a linear collider. The clean
LC environment allows a precise determination of the Higgs invisible
branching ratio over a broad range of Higgs masses (see {\emph{e.g.}}
Ref.~\cite{schumacher}).  We exemplarily show the improvement due the
measurement of ${\cal{J}}_1$ for the $m_{H_1}=155$ GeV scenario
discussed in Fig.~{\ref{fig:mud2}} of the previous section in
Fig.~\ref{fig:mud1b}. From Fig.~\ref{fig:mud2b} it also becomes clear
that the linear collider gives a good reconstruction of the Higgs
portal for percent level values of ${\cal{J}}$.

Due to the measurement of both ${\cal{J}}_1$ and ${\cal{R}}_1$ we can
reconstruct the intersection of both curves, yielding the mixing angle
\begin{equation}
  \cos^2\chi={\cal{J}}_1+{\cal{R}}_1\,.
\end{equation}
An independent measurement of $\sin^2\chi={\cal{J}}_2+{\cal{R}}_2$
overconstraints the system, giving rise to the sum rule
\begin{equation}
  {\cal{J}}_1+{\cal{R}}_1+{\cal{J}}_2+{\cal{R}}_2=1\,,
\end{equation}
which can be used to test the consistency of the portal model
Eq.~\gl{eq:potential} with experimental observations. We stress that
this is not possible at the LHC due to Eq.~\gl{eq:disentangle}.

\begin{figure}[!b]
  \begin{center}
    \includegraphics[width=0.55\textwidth]{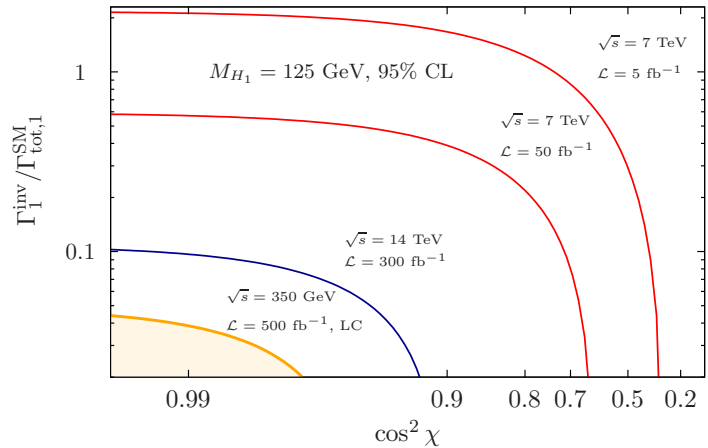}
    \caption{\label{fig5} 95\% CL contours for a measurement of
      $\Gamma_1^{\rm{hid}}/\Gamma_1^{\rm{SM}}$ at the LHC and a
      $350~\gev$ ILC. The LHC uncertainties are computed with
      {\sc{Sfitter}} \cite{sfitter} and the LC uncertainties are again
      adopted from Refs.~\cite{teslatdr}; the figure is taken from
      Ref.~\cite{Englert:2011aa}.}
  \end{center}
\end{figure}

Coming back to the strategy of approaching the SM with measurements
that constrain $\Gamma_1^{\rm{inv}}/\Gamma_1^{\rm{SM}}$, it is
worthwhile addressing the implications of the measured excess around
125~$\gev$ for the portal model. If this turns out to be the Higgs
then a measurement of $\Gamma_1^{\rm{inv}}/\Gamma_1^{\rm{SM}}$ give us
a measure of the compatibility of the experimental observations with
the SM.

Treating $\Gamma_1^{\rm{hid}}$ as a free parameter, we show 95 \%
confidence level contours in the
$\Gamma_1^{\rm{inv}}/\Gamma_1^{\rm{SM}}$-$\cos^2\chi$ plane in
Fig.~\ref{fig5}. The blue and red lines correspond to measurements at
the LHC, while the shaded are gives the prospects at a linear
collider. Obviously the current findings at the LHC are not good
enough from a statistical point of view to tell us wether or not we
observe the SM Higgs. These bounds improve when higher center of mass
energy and more integrated luminosity becomes available, but
systematic uncertainties saturate the LHC sensitivity at around
300~$\ifb$. 

A future linear collider has the potential to take this LHC legacy to
the next level: In Fig.~\ref{fig5} there is only a small region
untested for the LC curve. For the chosen set-up of $350~\gev,
500~\ifb$ the statistical and systematic uncertainties are comparable
\cite{schumacher}, hence further improvements can be expected by an
even larger data sample.

\subsection*{Acknowledgments}
We thank the organizers of the LC-Forum, in particular Gudrid
Moortgat-Pick, for the fruitful and friendly atmosphere during the
workshop. Support by the Durham Junior Research Fellowship scheme is
acknowledged.

\end{document}